\documentclass[aps,pre,amsmath,amssymb,superscriptaddress, footinbib]{revtex4-2}
\usepackage{graphicx}% Include figure files
\usepackage{color}
\usepackage{amsfonts}
\usepackage{physics}
\usepackage{braket}
\usepackage{multibib}
\usepackage[english]{babel}
\usepackage{bm}

\begin{document}
\title{Microrheology near jamming}
\author{Yusuke Hara}
\email{hara-yusuke729@g.ecc.u-tokyo.ac.jp}
\affiliation{Graduate School of Arts and Science, The University of Tokyo, Komaba, Tokyo 153-8902, Japan}
\author{Hideyuki Mizuno}
\affiliation{Graduate School of Arts and Science, The University of Tokyo, Komaba, Tokyo 153-8902, Japan}
\author{Atsushi Ikeda}
\affiliation{Graduate School of Arts and Science, The University of Tokyo, Komaba, Tokyo 153-8902, Japan}
\affiliation{Research Center for Complex Systems Biology, Universal Biology Institute, The University of Tokyo, Komaba, Tokyo 153-8902, Japan}

\date{\today}

\begin{abstract}
The jamming transition is a nonequilibrium critical phenomenon, which governs characteristic mechanical properties of jammed soft materials, such as pastes, emulsions, and granular matters.
Both experiments and theory of jammed soft materials have revealed that the complex modulus measured by conventional macrorheology exhibits a characteristic frequency dependence.
Microrheology is a new type of method to obtain the complex modulus, which transforms the microscopic motion of probes to the complex modulus through the generalized Stokes relation (GSR).
Although microrheology has been applied to jammed soft materials, its theoretical understanding is limited.
In particular, the validity of the GSR near the jamming transition is far from obvious since there is a diverging length scale $l_c$, which characterizes the heterogeneous response of jammed particles.
Here, we study the microrheology of jammed particles by theory and numerical simulation.
First, we develop a linear response formalism to calculate the response function of the probe particle, which is transformed to the complex modulus via the GSR.
Then, we apply our formalism to a numerical model of jammed particles and find that the storage and loss modulus follow characteristic scaling laws near the jamming transition.
Importantly, the observed scaling law coincides with that in macrorheology, which indicates that the GSR holds even near the jamming transition.
We rationalize this equivalence by asymptotic analysis of the obtained formalism and numerical analysis on the displacement field of jammed particles under a local perturbation.
\end{abstract}

%\keywords{}

\maketitle


\section{Introduction}

When gradually compressed, an assembly of particles acquires rigidity at a certain packing fraction.
This nonequilibrium critical phenomenon is the jamming transition that governs the properties of jammed soft materials, such as pastes, emulsions, and granular matter~\cite{van2009}. 
The simplest model of jammed soft materials is an assembly of frictionless spherical particles interacting via pairwise harmonic potentials. 
Importantly, various observables in this model show critical power-law behaviors near the jamming point~\cite{ohern2002, ohern2003, wyart2005a, wyart2005b, silbert2005, silbert2009, goodrich2012}.
The pressure $P$ of the system can be used as a control parameter and the jamming point corresponds to $P=0$. 
The linear shear modulus $G_0$ is proportional to $\sqrt{P}$~\cite{ohern2003, goodrich2012}.
The vibrational density of states (vDOS) $D(\omega)$ shows a characteristic plateau in frequency $\omega$ down to the onset frequency $\omega_*$~\cite{silbert2005, silbert2009}.
Approaching the jamming point, the onset frequency vanishes as $\omega_* \propto \sqrt{P}$.
Several mean-field theories, such as the effective medium theory and the replica theory, successfully reproduce these critical behaviors~\cite{wyart2005a, wyart2005b, wyart2010, yan2016, degiuli2014a, degiuli2014b, franz2015}.

The mechanical properties of soft matter are generally frequency-dependent, and the complex modulus $G^*(\omega)$ carries the relevant information on the linear mechanical response~\cite{larson1999}.
This quantity is defined through the stress response of the system under the oscillatory strain: the real part $G^{\prime} (\omega)$, called a storage modulus, represents the elastic response, while the imaginary part $G^{\prime \prime} (\omega)$, called a loss modulus, represents the viscous response.
The complex modulus has been studied for jammed particles. 
Experimental results suggest that $G^{\prime}(\omega)$ and $G^{\prime \prime}(\omega)$ are proportional to $\sqrt{\omega}$ in a certain frequency range~\cite{cohen1998, gopal2003, besson2008, krishan2010, kropka2010}.
This behavior was also observed in numerical simulation and derived theoretically~\cite{tighe2011, baumgarten2017}, which shows that this scaling $G^*(\omega) \propto \sqrt{i\omega}$ is a consequence of the characteristic plateau in the vibrational density of state $D(\omega)$.

Microrheology is a new type of measurement to evaluate the complex modulus~\cite{squires2010, furst2017book}.
In this measurement, the complex modulus is evaluated by probing the microscopic motion of particles suspended in a system. 
The key assumption is the so-called generalized Stokes relation (GSR), which relates the microscopic response function $\alpha(\omega)$ to the macroscopic complex modulus by $G^*(\omega) = 1/(3 \pi a_p \alpha(\omega))$, where $a_p$ is the diameter of the probe particle. 
In the simplest form of active microrheology, the response function $\alpha(\omega)$ is obtained by applying an oscillatory external force to the probe particle and measuring its displacement~\cite{gittes1997,schnurr1997,levine2000, levine2001,atakhorrami2006, mizuno2008, wilson2009}. 
In passive microrheology, one measures the spontaneous motion of particles under a thermal fluctuation, such as the mean-square displacement~\cite{mason1995, mason1997, mason2000}. 
The fluctuation-dissipation theorem then relates this fluctuation to the response function $\alpha(\omega)$. 
In both active and passive microrheology, one then uses the GSR to obtain the complex modulus.   
Microrheology has attracted attention for its ability to measure over a wider frequency range than conventional macrorheology, and its applicability to small systems where macrorheology is difficult to apply~\cite{furst2017book}.
Microrheology has been applied to jammed particles, such as emulsions~\cite{liu1996} and colloidal glasses~\cite{mason1995}. 
The experiments show that the strage and loss modulus measured in microrheology exhibit $G^{\prime}(\omega) \propto \sqrt{\omega}$ and $G^{\prime \prime}(\omega) \propto \sqrt{\omega}$ in a certain frequency range~\cite{liu1996,mason1995}, as in the case of macrorheology.
More recently, microrheology was applied to the cytoplasm, which is composed of jammed droplets of biopolymers, and a similar frequency dependence $G^*(\omega) \propto \sqrt{i\omega}$ was observed~\cite{nishizawa2017(a), nishizawa2017(b)}.

However, although these experimental results are enlightening, the theoretical background of the microrheology of jammed particles is unclear, especially in the vicinity of the jamming point. 
The key assumption in microrheology is the GSR, which assumes that the macroscopic constitutive equation remains valid down to the length scale of the probe particle.
%In general, when the size of the probe is smaller than the characteristic length scale of the system, the difference between micro- and macrorheology appears, and thus the GSR breaks down \cite{valentine2001, wong2004, tuteja2007}.
%\red{
However, generally speaking, in the case that the system has a characteristic length scale below which the continuum description of the system does not hold, and the size of the probe is smaller than this length scale, the difference between micro- and macrorheology appears, and thus, the GSR breaks down \cite{valentine2001, wong2004, tuteja2007}.%}
For jammed particles, there seems to be such a characteristic length scale~\cite{silbert2005,ikeda2013,lerner2014,karimi2015}. 
It is known that the displacement field of jammed particles under a local perturbation becomes highly heterogeneous and exhibits nonelastic behaviors~\cite{lerner2014}. 
This displacement field is quite disordered and no longer resembles what elastic theory predicts.
It becomes consistent with elastic theory only when the length scale is larger than the characteristic length $l_c$, which diverges as $l_c \propto P^{-1/4}$ at the jamming point~\cite{silbert2005,ikeda2013,lerner2014}.
This indicates the possibility that the macroscopic constitutive law does not work in the length scale of the probe particle, and then the GSR breaks down near the jamming point.

In this work, we theoretically study the microrheology of jammed particles.
In particular, we aim to construct a theoretical framework to calculate the complex modulus in microrheology, and clarify the relationship between the micro- and macrorheology of jammed particles in the vicinity of the jamming point.
We show that the two are indeed equivalent even very close to the jamming transition.
The paper is organized as follows.
In Section 2, we construct a theoretical framework to treat the microrheology of jammed particles.
Here, a linear response formalism is established to calculate the complex modulus in microrheology.
In Section 3, we apply this framework to a numerical model of jammed particles. 
We show that the complex modulus in microrheology is equivalent to that in macrorheology.
In Section 4, we discuss the equivalence between micro- and macrorheology from two different perspectives.
We first derive the scaling laws observed in Section 3, which explain the equivalence in terms of the nature of the vibrational modes.
We next analyze the displacement fields in jammed particles when a local force is applied. 
This analysis explains how the equivalence survives even when the characteristic length scale $l_c$ appears.

\section{Theoretical framework}

In this section, we develop a linear response formalism to calculate the complex modulus in the microrheology of jammed particles. 
We consider a $d$-dimensional system of $N-1$ constituent particles and a probe particle.
Hereafter, roman indices (e.g. $i, j$) denote the particles in the system that run from $1$ to $N$, and $p$ indicates the label of the probe particle. 
We assume that the system is initially in a mechanically stable state; in other words, the system is at a local energy minimum in the potential energy landscape.
We then consider the dynamics of the system when a weak oscillatory force is applied to the probe particle.

As we focus on the linear response of the system, we may expand the potential energy $U$ to the second-order in the particle's displacements.
The displacement vector of the $i$-th particle is denoted by $\vec{u}_i$ and its collection by a $dN$-dimensional vector $\ket{u} \equiv (\vec{u}_1^{\mathrm{T}}, \cdots, \vec{u}_N^{\mathrm{T}})^{\mathrm{T}}$.
%\red{
In this paper, we treat vectorial quantities as vertical vectors, and T denotes the transposition operator.%}
We also introduce the $dN \times dN$ dynamical matrix $\mathcal{M}$ whose $(i,j)$ block is given by 
\begin{equation}
   \mathcal{M}_{ij} = \pdv{U}{\vec{u}_j}{\vec{u}_i}.
\end{equation}
The harmonic potential  energy can then be written as 
\begin{align}
U = U_0 + \frac{1}{2} \expval{\bm{\mathcal{M}}}{u}, 
\end{align}
where $U_0$ is the potential energy at mechanical equilibrium. 
Because we are mainly interested in dense colloidal suspensions and concentrated emulsions, we focus on the overdamped limit of the equation of motion. 
With the harmonic approximation and the overdamped limit, a general equation of motions is given by
\begin{equation}
    \bm{C} \ket{\dot{u}} = - \mathcal{M} \ket{u} + \ket{F},
\end{equation}
where $\bm{C}$ is a damping coefficient matrix, whose explicit form depends on the details of the dissipation in the system, but here, we only assume it to be regular and symmetric. 
Note that $\bm{C}$ is diagonal for the so-called Stokes drag model, while $\bm{C}$ has a slightly more complicated form for the contact damping model~\cite{durian1995, durian1997}.
$\ket{F}$ is the external force applied to particles.
We apply an external force $F_{p, \alpha}$ to the probe particle in the $\alpha$ direction, and then $\ket{F} = (\vec{0}^{\mathrm{T}}, \cdots, F_{p, \alpha} , \cdots, \vec{0}^{\mathrm{T}})^{\mathrm{T}}$.
By the Fourier transformation, the equation of motions can be recast into the following:
\begin{equation}
    (\mathcal{M} + i \omega\bm{C} ) \ket{\hat{u}(\omega)} = \ket{\hat{F}(\omega)}, \label{eq:eq-motion}
\end{equation}
or equivalently, 
\begin{equation}
    \ket{\hat{u}(\omega)} = \bm{G}(\omega) \ket{\hat{F}(\omega)}, 
\end{equation}
where $\bm{G}(\omega) \equiv (\mathcal{M} + i \omega \bm{C})^{-1}$ is the Green's function.
Each element of the Green's function $G_{i\alpha, j\beta}(\omega)$ expresses the displacement of the $i$-th particle in the $\alpha$ direction caused by a unit force imposed on the $j$-th particle in the $\beta$ direction.
Therefore, the response of the probe particle itself is $G_{p\alpha, p\alpha}(\omega)$, which corresponds to the response function $\alpha(\omega)$ measured in microrheology.

To invert $\left(\mathcal{M} + i \omega \bm{C}\right)$, it is useful to study the following generalized eigenvalue problem:
\begin{equation}
    \omega_k^2 \bm{C} \ket{e(\omega_k)} = \mathcal{M} \ket{e(\omega_k)} \label{eq:general-eigen},
\end{equation}
where $\omega_k$ is a generalized eigenfrequency and $\ket{e(\omega_k)}$ is a generalized eigenvector. 
We multiply $\bm{C}^{-\frac{1}{2}}$ on both sides of this equation to obtain the following:
\begin{equation}
    \omega_k^2 \bm{C}^{\frac{1}{2}} \ket{e(\omega_k)} = \bm{C}^{-\frac{1}{2}} \mathcal{M} \bm{C}^{-\frac{1}{2}} \bm{C}^{\frac{1}{2}} \ket{e(\omega_k)}.
\end{equation}
Note that $\bm{C}^{\frac{1}{2}}$ is symmetric as $\bm{C}$ is symmetric. 
Then, by introducing $\ket{\tilde{e}(\omega_k)} = \bm{C}^{\frac{1}{2}} \ket{e(\omega_k)}$ and $\tilde{\mathcal{M}} = \bm{C}^{ - \frac{1}{2}} \mathcal{M} \bm{C}^{ - \frac{1}{2}}$, Eq.~(\ref{eq:general-eigen}) becomes
\begin{equation}
    \omega_k^2 \ket{\tilde{e}(\omega_k)} = \tilde{\mathcal{M}} \ket{\tilde{e}(\omega_k)}.
\end{equation}
Therefore, $\omega_k$ and $\ket{\tilde{e}(\omega_k)}$ are a eigenfrequency and eigenvector of the matrix $\tilde{\mathcal{M}}$.

Now we can express the Green's function using $\omega_k$ and $\ket{\tilde{e}(\omega_k)}$. 
We introduce $\tilde{\bm{E}} = (\ket{\tilde{e}(\omega_1)}, \ket{\tilde{e}(\omega_2)}, \cdots, \ket{\tilde{e}(\omega_{dN})})$ and define $\bm{E} = \bm{C}^{-\frac{1}{2}} \tilde{\bm{E}}$.
Then Eq.~(\ref{eq:eq-motion}) can be recast as:
\begin{equation}
	\bm{E}^{\mathrm{T}}\left( \mathcal{M} + i \omega \bm{C} \right) \bm{E} \bm{E}^{-1} \ket{\hat{u}(\omega)}  = \bm{E}^{\mathrm{T}}\ket{\hat{F}(\omega)}.
\end{equation}
By using the identities $\bm{E}^{\mathrm{T}} \bm{C} \bm{E} = \bm{I}$ and  $\bm{E}^{\mathrm{T}} \mathcal{M} \bm{E} = \text{diag} \left(\omega_1^2, \cdots, \omega_{dN}^2 \right) \equiv \bm{\Omega}$, this equation can be solved as 
\begin{equation}
    \ket{\hat{u}(\omega)}  = \bm{E} \left( \bm{\Omega} + i \omega \bm{I} \right)^{-1} \bm{E}^{\mathrm{T}}\ket{\hat{F}(\omega)}.
\end{equation}
Therefore, the Green's function $\bm{G}(\omega)$ can be expressed in terms of the eigenfrequencies and eigenvectors of the matrix $\tilde{\mathcal{M}}$ as: 
\begin{align}
	G_{i\alpha, j\beta}(\omega) 
	& = \sum_{m, n, \gamma, \delta} (\bm{C^{-\frac{1}{2}}})_{i\alpha, m\gamma} \sum_{k} \frac{\tilde{e}_{m, \gamma}(\omega_k) \tilde{e}_{n, \delta}(\omega_k) }{\omega_k^2 + i \omega} (\bm{C^{-\frac{1}{2}}})_{n\delta, j\beta}. 
	%\mbox{\red{summarion}}
\end{align}
By setting $i=j=p$, we obtain $G_{p\alpha, p\beta}(\omega)$, the response function in microrheology. 
Inserting this into the generalized Stokes relation, we obtain the complex modulus measured in microrheology as 
\begin{align}
	G^*_{m} (\omega) 
	& = \frac{1}{3 \pi a_p G_{p\alpha, p\alpha}(\omega)} \notag \\
	& = \frac{1}{3 \pi a_p} \frac{1}{\sum_{m, n, \gamma, \delta}(\bm{C^{-\frac{1}{2}}})_{p\alpha, m\gamma} \sum_{k} \frac{\tilde{e}_{m, \gamma}(\omega_k) \tilde{e}_{n, \delta}(\omega_k) }{\omega_k^2 + i \omega} (\bm{C^{-\frac{1}{2}}})_{n\delta, p\alpha}}, 
	%\mbox{\red{summarion}}
	\label{eq:complexmodulus}
\end{align}
where $a_p$ denotes the diameter of the probe particle. 
Note that the subscript $m$ for $G^*_{m} (\omega)$ indicates the microrheology.

\section{Microrheology of jammed particles}

In this section, we apply the formalism developed above to a numerical model of jammed particles to calculate the complex modulus in microrheology. 

\subsection{Setup}

We consider a $d=3$ system of $N-1$ monodisperse spherical particles interacting through the two-body potential $v(r)$ and a probe particle that interacts with the other particle through the two-body potential $v_p(r)$. 
The potential is set to be $v(r_{ij}) = \frac{\epsilon}{2} (a-r_{ij})^2 \Theta(a-r_{ij})$ and $v_p(r_{pj}) = \frac{\epsilon_p}{2} (\frac{a_p+a}{2}-r_{pj})^2 \Theta(\frac{a_p+a}{2}-r_{pj})$, where $r_{ij}$ and $r_{pj}$ are the interparticle distance. 
The constants $\epsilon$ and $a$ express the hardness and diameter of the constituent particles, respectively, and $\epsilon_p$ and $a_p$ are those of the probe particle.
The dissipation in the system is modeled by the Stokes drag model as $C_{i \alpha, j \beta} = C_0 \delta_{ij} \delta_{\alpha \beta}$, except for $C_{p \alpha, p \beta} = C_p \delta_{\alpha \beta}$ for the probe particle. 
Here, we focus on the case in which the probe particle and the constituent particles are identical: $a_p = a$, $\epsilon_p = \epsilon$, and $C_p = C_0$. 
In this case, the system is completely monodisperse, and an arbitrarily chosen particle (with the index $p$) can be regarded as the probe particle. 
We checked that the size difference between the probe and constituent particles does not change our results much.
For completeness, we present the result for $a_p = 4a$ in Appendix A. 
The units of the length, energy, and time are set to be $a$, $a^2 \epsilon$, and $C_0 / \epsilon$ respectively.
The total number of particles is $N=8000$.
We study the mechanical equilibrium packings with different pressures ranging from $P =10^{-2}$ to $10^{-7}$.
We generate such packings by the iterative compression and decompression algorithm followed by energy minimization \cite{bitzek2006, guenole2020}.

In this work, we focus on the \textit{unstressed} system, in which the normal force is set to zero in the mechanical equilibrium packing. 
This approximation makes it easier to analyze the critical behavior near the jamming transition~\cite{tighe2011}. 
For the unstressed system, the $(i,j)$ block of the dynamical matrix is given by  
\begin{align}
    \mathcal{M}_{ij} 
    = - \delta_{\langle ij \rangle} \hat{r}_{ij} \hat{r}_{ij}^{\mathrm{T}} + \delta_{ij} \sum_{j' \in \partial i} \hat{r}_{ij'} \hat{r}_{ij'}^{\mathrm{T}},
\end{align}
where $\delta_{\langle ij \rangle}=1$ when particles $i$ and $j$ are in contact, $\partial i$ represents the set of particles in contact with particle $i$, and $\hat{r}_{ij} = \vec{r}_{ij}/r_{ij}$ where $\vec{r}_{ij} = \vec{r}_j - \vec{r}_i$. 
Because the damping coefficient matrix is a unit matrix, the Green's function can be written as 
\begin{equation}
    G_{p\alpha, p\alpha}(\omega) = \sum_{k} \frac{ \abs{\tilde{e}_{p, \alpha}(\omega_k)}^2 }{\omega_k^2 + i \omega}. 
\end{equation}
Note that $\omega_k^2$ and $\tilde{e}_{i, \alpha}$ are identical to the eigenvalues and eigenvectors of $\mathcal{M}$ because $\tilde{\mathcal{M}}$ is identical to $\mathcal{M}$, as the damping coefficient matrix is a unit matrix.

We can now calculate the complex modulus $G^*_m(\omega)$ using $G_{p\alpha, p\alpha}(\omega)$ via Eq.~(\ref{eq:complexmodulus}). 
Since the probe particle is identical to the other constituent particles, we can regard each of the constituent particles as the probe particle. 
This makes it possible to take an average over the choice of the probe particle to improve our statistics. 
In practice, we consider two different procedures.
One way is 
\begin{align}
    G^*_{m(G)}(\omega)
    & \equiv \expval{\frac{1}{3 \pi G_{p \alpha, p\alpha}(\omega)}} \notag \\
    & = \frac{1}{3N} \sum_{i, \alpha} \frac{1}{3 \pi G_{i \alpha, i\alpha}(\omega)} \notag \\
    & = \frac{1}{9 \pi N} \sum_{i, \alpha} \frac{1}{\sum_{k} \frac{ \abs{\tilde{e}_{i, \alpha}(\omega_k)}^2 }{\omega_k^2 + i \omega}},  \label{eq:ave-mod-G}
\end{align}
where the complex modulus for each probe particle is calculated and then averaged. 
This averaging procedure is indicated by the subscript $(G)$. 
The other is 
\begin{align}
    G^*_{m(\alpha)}(\omega)
    & \equiv \frac{1}{3 \pi \expval{G_{p \alpha, p\alpha}(\omega)}} \notag \\
    & = \frac{1}{\frac{\pi}{N} \sum_{i, \alpha} G_{i \alpha, i\alpha}(\omega)} \notag \\
    & = \frac{N}{\pi} \frac{1}{\sum_{k} \frac{1}{\omega_k^2 + i \omega}}, \label{eq:LR-modulus}
\end{align}
where the response functions for each probe particle are averaged. 
This averaging procedure is indicated by the subscript $(\alpha)$. 
In this case, the trace of the Green's function is taken first, and then the eigenvectors do not appear in the final expression. 
These two ways of averaging procedures have experimental counterparts. 
For example in passive microrheology, the former corresponds to the situation in which we first calculate the complex modulus for each probe particle using the MSD of each probe particle, and then take an average. 
In the latter case, we first take an average of the MSD of the probe particles, and then invert it to obtain the complex modulus.
In the following, we show that these two averaging procedures give essentially the same results for jammed particles.

\subsection{Results}

\begin{figure}[h]
\centering
    \includegraphics[width=0.4\linewidth]{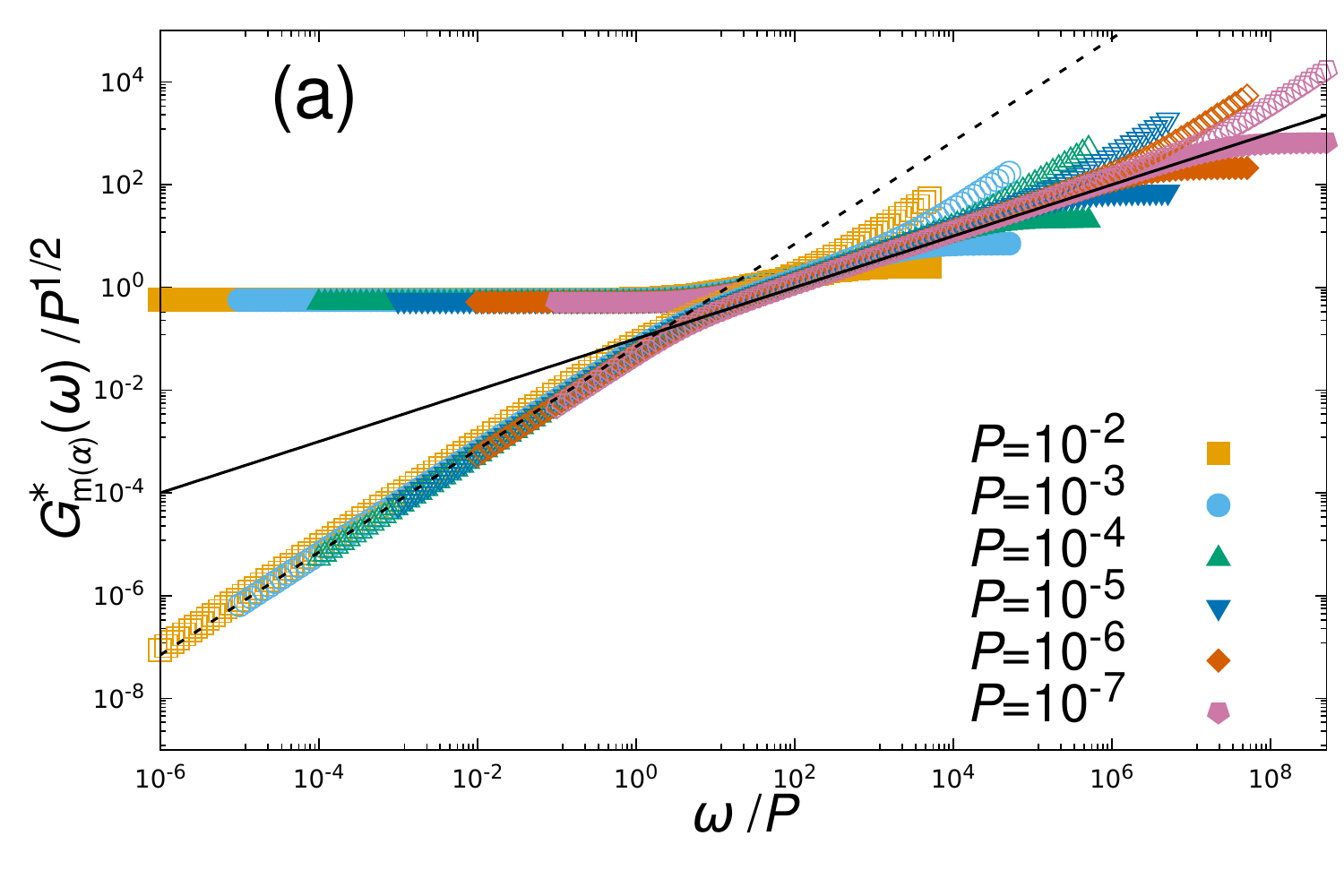}
    \includegraphics[width=0.4\linewidth]{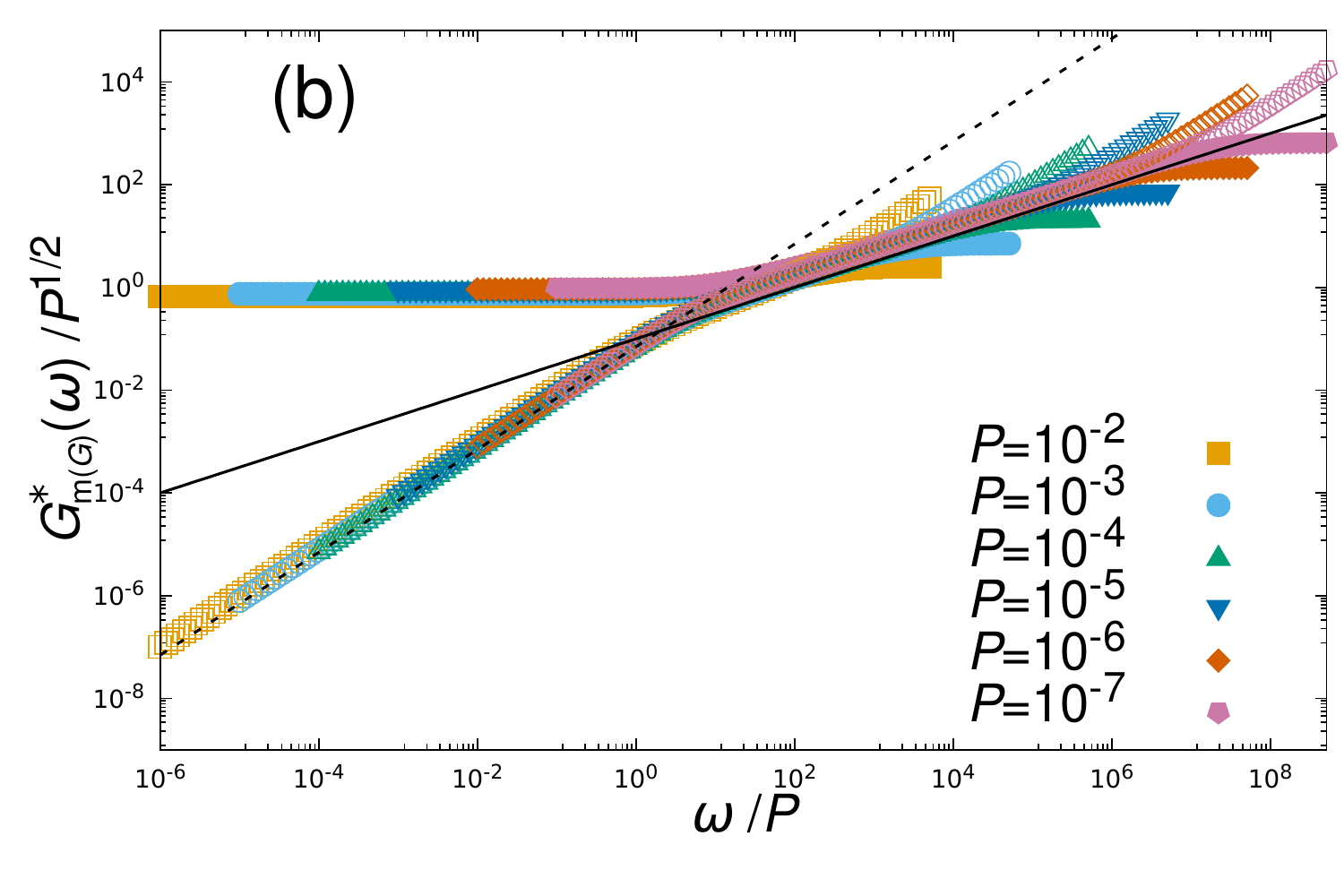}
    \includegraphics[width=0.4\linewidth]{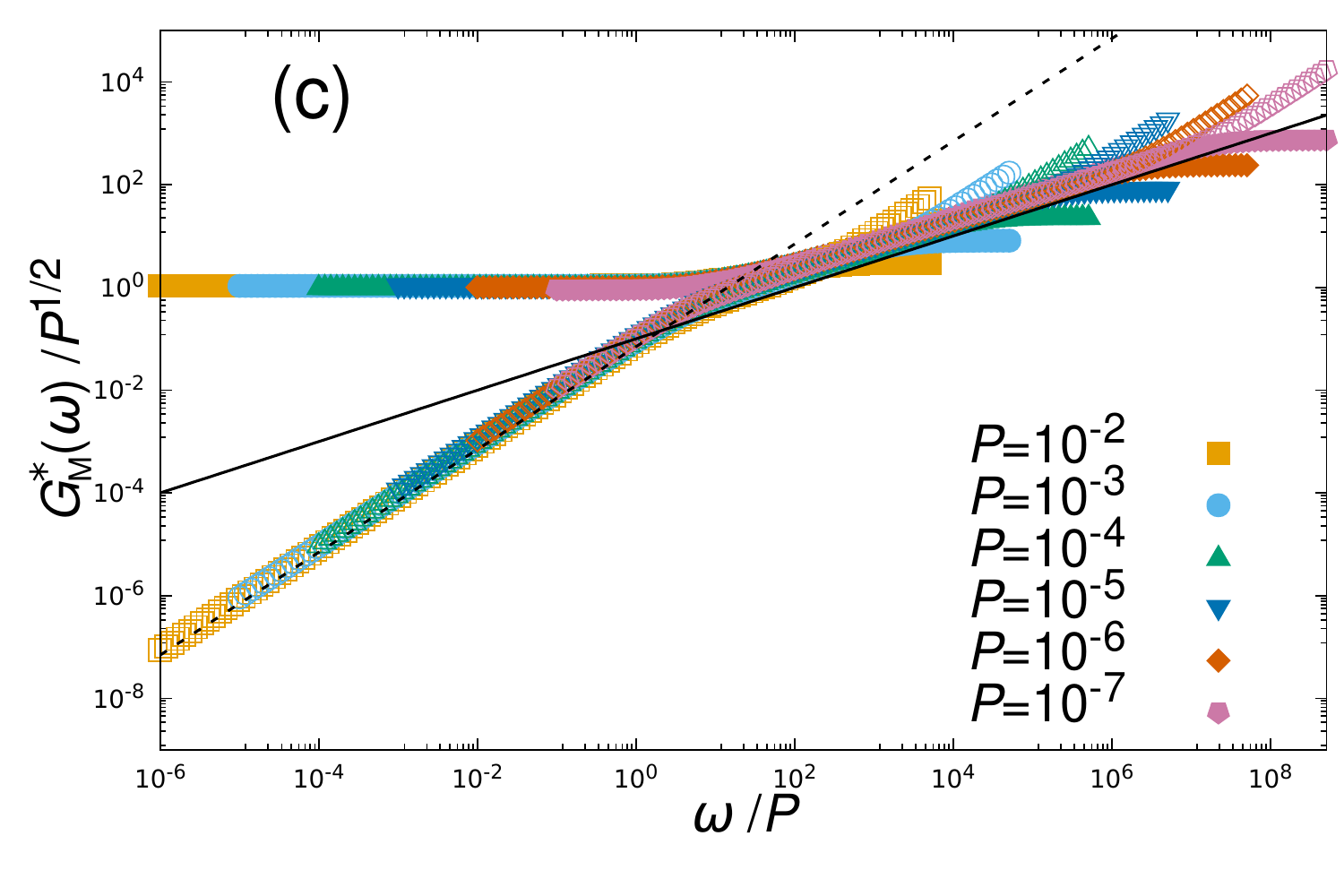}
    \includegraphics[width=0.4\linewidth]{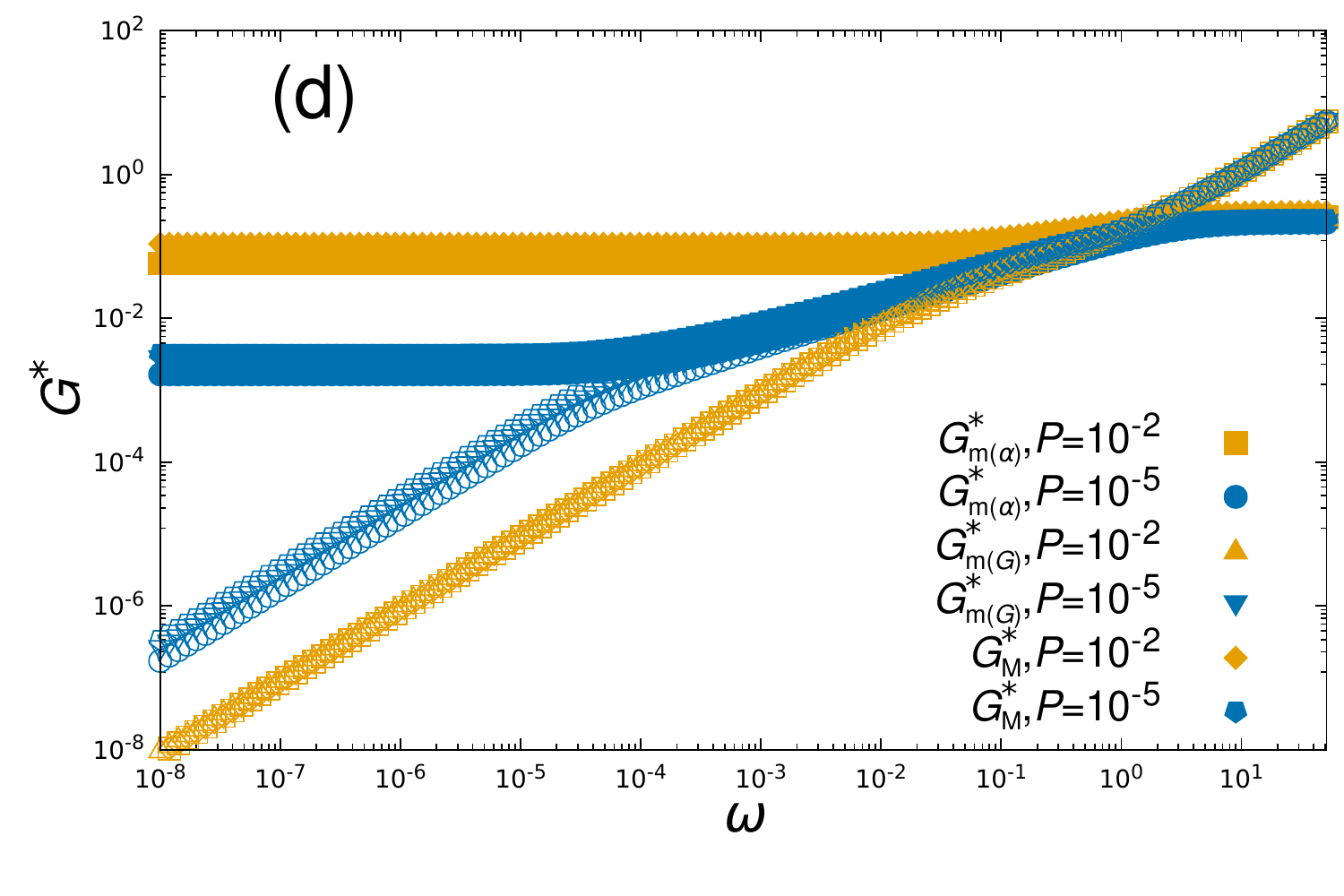}
    \caption{
	The storage modulus (filled symbols) and loss modulus (open symbols) calculated via (a,b) microrheology and (c) macrorheology are presented.
	We prepare 20 independent packings and the moduli are averaged over these different samples.
    The solid and dashed lines represent $\sqrt{\omega}$ and $\omega$ scaling respectively.
	The complex modulus is rescaled by $P^{\frac{1}{2}}$, and the frequency is rescaled by $P$.
	All of them show a very good collapse with rescaled variables.
    (d) The complex moduli of micro- and macrorheology at $P=10^{-2}, 10^{-5}$ are presented in the same plot.
    This plot shows that not only the scaling but also the magnitudes of the modulus are almost the same in the different measurements.
    }
\end{figure}

Fig.~1~(a) and (b) are the scaling plots of the real and imaginary parts of the obtained complex modulus of microrheology $G^*_{m(\alpha)}(\omega)$ and $G^*_{m(G)}(\omega)$. 
In both cases, the frequency is scaled by the pressure $P$ and the modulus by $P^{1/2}$ so that the results at different $P$ collapse very well. 
At higher frequency $\omega \gtrsim P$, the real and imaginary parts of various $P$ collapse onto the same curve $(\omega/P)^{1/2}$. 
This means that the complex modulus can be expressed as $G^*_m(\omega) \sim (i \omega)^{1/2}$ irrespective of $P$ in this frequency range. 
At lower frequency $\omega \lesssim P$, the real part quickly approaches a constant, and then the static limit of the real part becomes $G'_m(0) \propto P^{1/2}$. 
Likewise, the imaginary part quickly approaches $G''_m(\omega)/P^{1/2} \propto \omega/P$ and then $G''_m(\omega) \propto \eta_m \omega$ with the viscosity diverging as $\eta_m \propto P^{-1/2}$. 
Interestingly, these observations apply for both $G^*_{m(\alpha)}$ and $G^*_{m(G)}$. 
This tells us that the averaging procedure does not matter for the microrheology of jammed particles.

Broadly, the results presented above are similar to those of macrorheology~\cite{tighe2011, baumgarten2017}.
To discuss this point clearly, we also computed the complex modulus of macrorheology $G^*_M(\omega)$ in the same setting with the same configurations.  
The details of the computations are shown in Appendix B. 
Fig.~1~(c) shows the scaling plot of $G^*_M(\omega)$. 
Clearly, $G^*_M(\omega)$ and $G^*_m(\omega)$ follow the same scaling laws. 
Moreover, if we compare their numerical values more closely, then we find that they are almost equivalent. 
%\textcolor{red}{It would be good to plot all the results in the same panel and show it as (d)}
This equivalence is directly addressed in Fig.~1~(d), where we plot three different complex modulus  $G^*_{m(\alpha)}$, $G^*_{m(G)}$ and $G^*_M(\omega)$ at two different pressures $P=10^{-2}$ and $10^{-5}$, without scaling by $P$. 
Clearly, the three different complex moduli are almost the same. 

Therefore, these results establish the equivalence between the micro and macrorheology for jammed particles.
Such equivalence is known to appear in systems where the macroscopic constitutive law is valid down to the scale of the probe particle \cite{furst2017book}. 
A trivial example is a Newtonian fluid \cite{valentine2001}. 
However, this explanation seems not to work for jammed particles as the description of continuum elasticity is known to breakdown near the jamming transition~\cite{lerner2014}.
This suggests that the mechanism behind the equivalence in jammed particles would be different from that in a Newtonian fluid. 
In the following section, we discuss this issue.

\section{Discussion}

In this section, we discuss the reason why microrheology and macrorheology give the essentially same results, from two different perspectives.
First, we perform the asymptotic analysis of Eqs.~(\ref{eq:ave-mod-G}) and (\ref{eq:LR-modulus}), and derive the scaling laws observed in the previous section. 
This analysis explains the equivalence between micro- and macrorheology in terms of the vibrational properties of jammed particles. 
Second, we analyze the displacement field of particles when applying a local force.
We show that elastic theory description breaks down in the length scale $1 \lesssim r \lesssim l_c$, as has been discussed~\cite{lerner2014}, but at the particle scale $r \approx 1$, the amplitude of particle displacement can still be described by the macroscopic elastic modulus.

\subsection{Scaling analysis}

We first derive the scaling laws for $G^*_{m(\alpha)}(\omega)$ in microrheology. 
The derivation goes in a very similar way as for $G^*_{M}(\omega)$  in macrorheology~\cite{tighe2011}. 
From Eq.~(\ref{eq:LR-modulus}), $G^*_{m(\alpha)}(\omega)$ can be expressed by the vibrational density of states $D(\omega) = \frac{1}{3N} \sum_k \delta(\omega - \omega_k)$ as 
\begin{equation}
    \frac{1}{G^*_{m(\alpha)}(\omega)} =
	3 \pi \int d\omega^{\prime} \frac{D(\omega^{\prime})}{(\omega^{\prime})^2 + i\omega}. 
\end{equation} 
Hereafter, we assume that the system is near jamming $P \ll 1$, and we focus on the low frequency regime $\omega \ll 1$. 
In the {\it unstressed} system under this condition, $D(\omega)$ is known to have both phonon and anomalous parts:
\begin{align}
	D(\omega) \sim 
	\begin{cases}
		A_D \omega^{d-1} & (\omega < \omega_{*}) \\
        	\omega^0 & (\omega > \omega_{*}) 
 	\end{cases}
\end{align}
where $\omega_{*}$ is the onset frequency of the anomalous part and proportional to $\sqrt{P}$. 
$A_D \propto \omega_{*}^{-\frac{3}{2}}$ is the Debye level.

We first evaluate the contribution from the anomalous part. 
Note that we omit unimportant numerical factors in the following calculations.
By introducing $x = \omega^{\prime}/\sqrt{\omega}$, the integral can be rewritten as
\begin{align}
\int_{\omega_{*}}^{1} \frac{d\omega^{\prime}}{\left(\omega^{\prime}\right)^2 + i \omega} 
= \frac{1}{\sqrt{\omega}} \int^{\infty}_{\omega_{*}/\sqrt{\omega}} dx \frac{x^2 -i}{x^4 + 1}. 
\end{align}
%where the upper limit of the integral is replaced with $\infty$ since $\omega \ll 1$. 
In the high frequency regime $\omega_{*} \ll \sqrt{\omega}$, the integral can be expanded with respect to $\omega_{*}/\sqrt{\omega} \ll 1$, and the leading order term is 
\begin{align}
\frac{1}{\sqrt{\omega}} - i\frac{1}{\sqrt{\omega}}. 
\label{eq:plateau-high}
\end{align}
In the low frequency regime $\omega_{*} \gg \sqrt{\omega}$, the integral can be expanded with respect to $\sqrt{\omega}/\omega_{*} \ll 1$, and the leading order term is 
\begin{align}
\frac{1}{\omega_{*}} - i\frac{\omega}{\omega_{*}^3}. 
\label{eq:plateau-low}
\end{align}
Second, the contribution from the phonon part can be rewritten as 
\begin{align}
A_D \int_{0}^{\omega_{*}} d\omega^{\prime} \frac{(\omega^{\prime})^2}{\left(\omega^{\prime}\right)^2 + i \omega} 
= A_D \sqrt{\omega} \int_{0}^{\omega_{*}/\sqrt{\omega}} \frac{x^4 - i x^2}{x^4 + 1} dx 
\end{align}
In $\omega_{*} \ll \sqrt{\omega}$, this integral can be expanded with respect to $\omega_{\star}/\sqrt{\omega} \ll 1$, and the leading order terms are proportional to $\omega_{*}^{7/2} /\omega^2$ for the real part and $\omega_{*}^{3/2} /\omega$ for the imaginary part.  
Because these contributions are much smaller than Eq.~(\ref{eq:plateau-high}) in the focused frequency regime, we can ignore the phonon contribution in this regime. 
In $\omega_{*} \gg \sqrt{\omega}$, the integral can be expanded with respect to $\sqrt{\omega}/\omega_{*} \ll 1$, and the leading order term is 
\begin{align}
\frac{1}{\sqrt{\omega_{*}}} - i \sqrt{\frac{\omega}{\omega_{*}^3}}. 
\label{eq:phonon-low}
\end{align}
By collecting all the leading order terms and inverting them, we obtain the viscoelastic modulus as 
\begin{equation}
G^*_{m(\alpha)}(\omega) \approx
    \begin{cases}
\sqrt{\omega} + i\sqrt{\omega} & \left(\omega_{*}^2 \ll \omega \ll 1 \right) \\
\omega_{*} + i\frac{\omega}{\omega_{*}} & \left(\omega_{*}^3 \ll \omega \ll \omega_{*}^2 \right) \\
\omega_{*} + i\sqrt{\omega_{*} \omega} & \left(\omega \ll \omega_{*}^3 \right),
    \end{cases}
\end{equation}
%\red{
or equivalently,
\begin{equation}
G^*_{m(\alpha)}(\omega)/\omega_* \approx
    \begin{cases}
\sqrt{\omega/\omega_*} + i\sqrt{\omega} & \left(1 \ll \omega/\omega_* \ll 1/\omega_* \right) \\
1 + i\frac{\omega}{\omega_{*}^2} & \left(\omega_{*} \ll \omega/\omega_*^2 \ll 1 \right) \\
1 + i\sqrt{\omega/\omega_*} & \left(\omega/\omega_*^2 \ll \omega_{*} \right).
    \end{cases}
\end{equation}
%}
This scaling law is the same as that observed in the previous section (remind that $\omega_*\propto P^{1/2}$). 
Note that the lowest frequency behavior at $\omega \ll \omega_{*}^3$, which is the contribution from the phonon part, was not observed in our numerics. 
This is presumably due to the finite size effect. 

The derived scaling law for $G^*_{m(\alpha)}(\omega)$ is the same as that for $G^*_{M}(\omega)$ in macrorheology, apart from the lowest frequency behavior~\cite{tighe2011}. 
Comparing the derivation for $G^*_{m(\alpha)}(\omega)$ in here and $G^*_M(\omega)$ first derived in \cite{tighe2011} and summarized in Appendix B, we can see that the equivalence comes from the fact that all the anomalous vibrational modes share similar values for the term $\abs{\ip{s_n}{\gamma}}^2$ (see Appendix B for the notation), which means that all the anomalous vibrational modes equally overlap with the macroscopic shear motion of the system.
This property originates from the disordered and extended characteristics of the anomalous vibrational modes~\cite{wyart2005a, wilson2009, silbert2009}.
Therefore, these nature of the vibrational modes bring the equivalence between the micro- and macrorheology near the jamming.

In addition, we discuss the reason why $G^*_{m(G)}(\omega)$ follows the same scaling law as $G^*_{m(\alpha)}(\omega)$. 
To calculate $G^*_{m(G)}(\omega)$, we need to calculate $\abs{\tilde{e}_{p, \alpha}(\omega_k)}^2$. 
However, due to the disordered and extended nature of the vibrational modes, this term can be estimated as $\abs{\tilde{e}_{p, \alpha}(\omega_k)}^2 \approx \frac{1}{3N}$ independent of $p, \alpha$ and $k$.
By inserting this into Eq.~(\ref{eq:LR-modulus}), we obtain
\begin{align}
     G^*_{m(G)}(\omega)
    \approx
    \frac{N}{\pi} \frac{1}{\sum_{k} \frac{1}{\omega_k^2 + i \omega}} = G^*_{m(\alpha)}(\omega). 
\end{align}
Therefore, the equivalence between  $G^*_{m(\alpha)}(\omega)$ and  $G^*_{m(G)}(\omega)$ is also the consequence of the disordered and extended nature of the vibrational modes near the jamming point.

In summary, the analysis in this section shows that the disordered and extended nature of the vibrational modes in the anomalous part bring the equivalence among $G^*_{m(\alpha)}(\omega)$, $G^*_{m(G)}(\omega)$, and $G^*_{M}(\omega)$.

\subsection{Displacement field and microrheology}

The results and discussion in the previous sections are somewhat counterintuitive because it has been known that the response of jammed particles to a local perturbation is spatially heterogeneous. 
Namely, there is a diverging length scale $l_c$ in jammed particles, below which the displacement fields are disordered and cannot be described by elastic theory. 
Since microrheology measures the response in a local scale, it is unclear how the equivalence between the microrheology and macrorheology survives even with such nonelastic behavior. 
In this section, we discuss this point.

To this end, we analyze the displacement field induced by a point force. 
The displacement field can be decomposed into the ``diagonal'' part which describes the displacements of particles along the direction of the point force, and the ``off-diagonal'' part which is perpendicular to the point force.
For these contributions, we introduce the intensity of the displacement field $V(r)$.
Using the static limit $\omega \to 0$ of the Green's function, these quantities can be defined as follows:
\begin{align}
    & V_D (r) = \expval{\left( G_{ix, jx}(0) \right)^2 + \left( G_{iy, jy}(0) \right)^2 + \left( G_{iz, jz}(0) \right)^2}_r \\
    & V_O (r) = 2 \expval{\left( G_{ix, jy}(0) \right)^2 + \left( G_{iy, jz}(0) \right)^2 + \left( G_{iz, jx}(0) \right)^2}_r
\end{align}
where $\expval{\cdots}_r$ represents the average over the pairs of particles $(i,j)$ with $r_{ij}=r$.
We numerically calculate these quantities for the model discussed in the previous section. 
To observe the large scale behavior properly, we perform this calculation for a larger system with $N=2048000$. 
The details of the calculations are summarized in Appendix C. 
We also calculate these quantities within elastic theory, which we denote $V_{D, \text{el}}(r)$ and $V_{O, \text{el}}(r)$. 
Their explicit expressions are summarized in Appendix D. 
By comparing $V(r)$ and $V_{\text el}(r)$, we can discuss the nonelastic nature of the response of jammed particles. 

\begin{figure}[h]
\centering
    \includegraphics[width=0.4\linewidth]{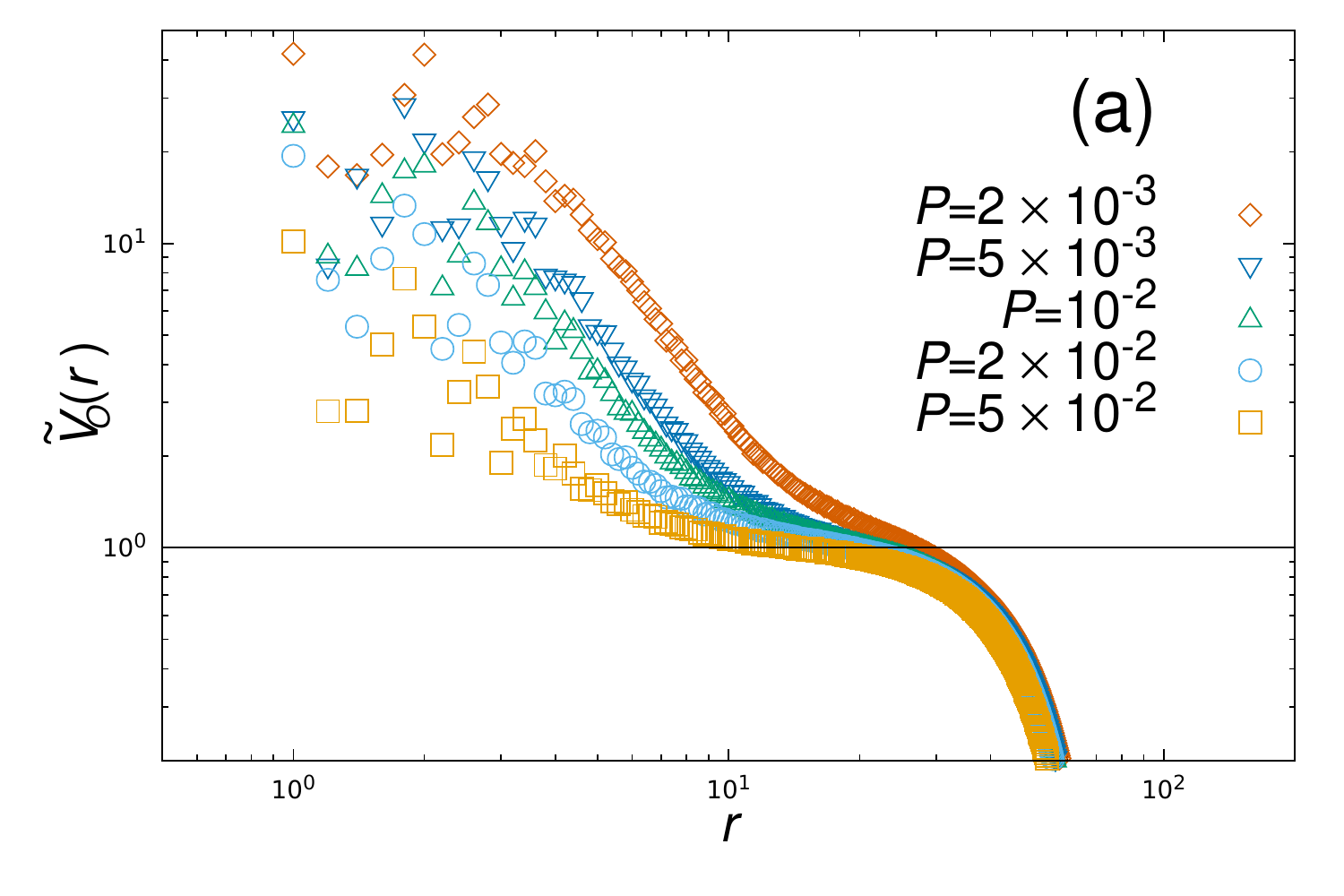}
    \includegraphics[width=0.4\linewidth]{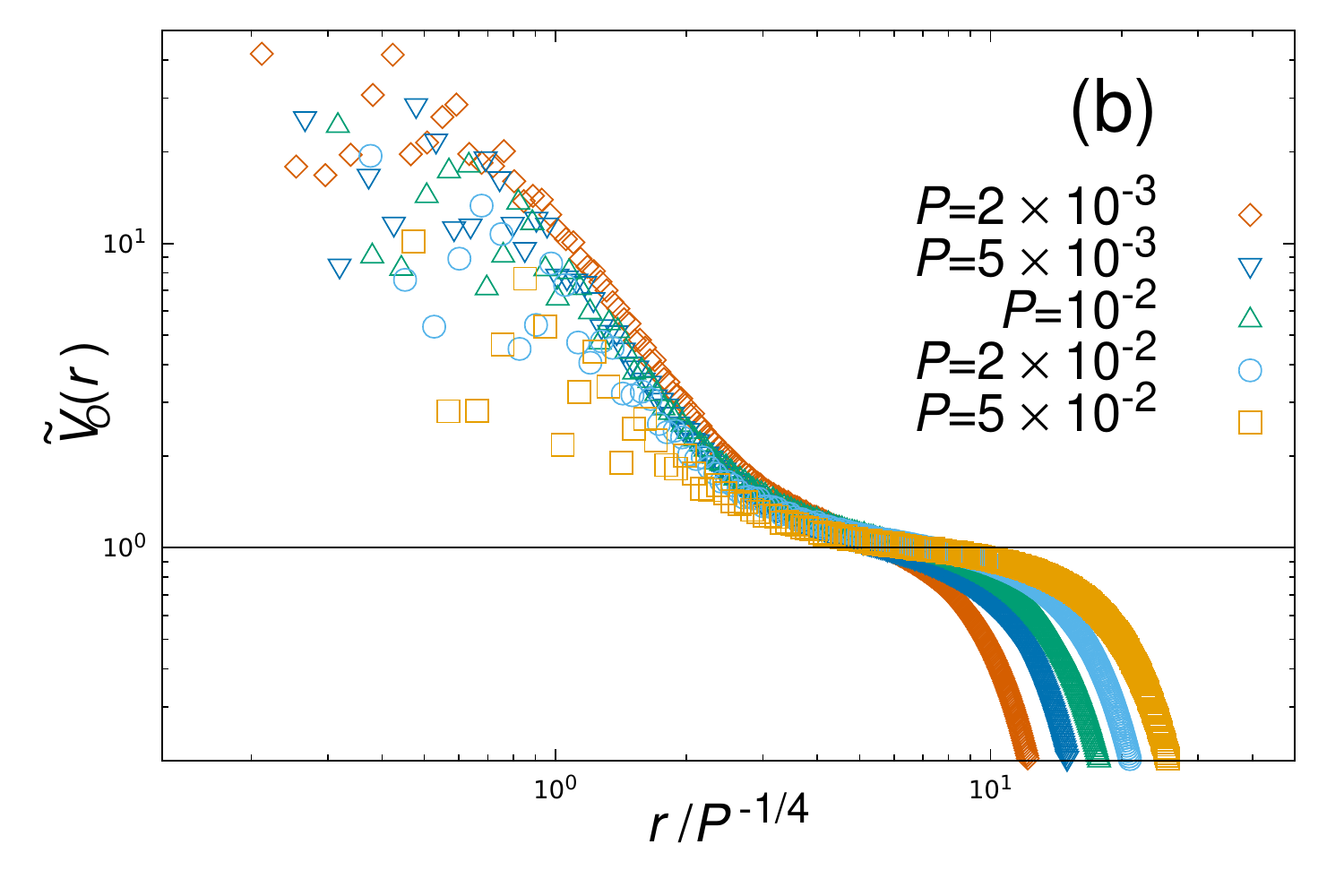}
    \caption{
    (a) The off-diagonal part of the displacement field $\tilde{V}_O(r) = V_O(r)/V_{O, \text{el}}(r)$, where $V_O(r)$ is the numerically obtained result and $V_{O, \text{el}}(r)$ is the prediction of elastic theory.
    For small $r$, $\tilde{V}_O(r)$ is larger than the prediction of elastic theory.
    For large $r$, its value converges to 1, which indicates the validity of elastic theory at a certain length scale $l_c$.
    Note that because of the finite size effect, $\tilde{V}_O(r)$ becomes significantly smaller than the prediction of elastic theory for $r \gg l_c$.
    (b) Same as (a), but the horizontal axis is rescaled as $r / P^{-1/4}$.
    The collapse confirms the presence of the diverging length scale $l_c \propto P^{-1/4}$ above which elastic theory works.
    }
    \label{fgr2}
\end{figure}

In Fig.~\ref{fgr2}~(a), we present $\tilde{V}_O(r) = V_O(r)/V_{O, \text{el}}(r)$, which should be 1 when elastic theory works. 
In all cases, $\tilde{V}_O(r)$ becomes larger than 1 when $r$ is small. 
This indicates the breakdown of elastic theory near the point force. 
As $r$ increases however, $\tilde{V}_O(r)$ approaches 1, thus elastic theory works well at a larger length scale.  
The characteristic length of this nonelastic behavior clearly increases as $P \to 0$. 
To quantify this, we plot $\tilde{V}_O(r)$ against $r/P^{-1/4}$ in Fig \ref{fgr2}~(b).
The rescaled datasets show a good collapse in the region of $\tilde{V}_O(r) \approx 1$. 
This shows that the length scale $l_c \propto P^{-1/4}$ controls the response.
Note that we cannot observe such behaviors for the diagonal part $\tilde{V}_D(r)$ in our simulations, which is presumably due to the finite size effect (see Appendix E).

\begin{figure}[h]
    \includegraphics[width=0.4\linewidth]{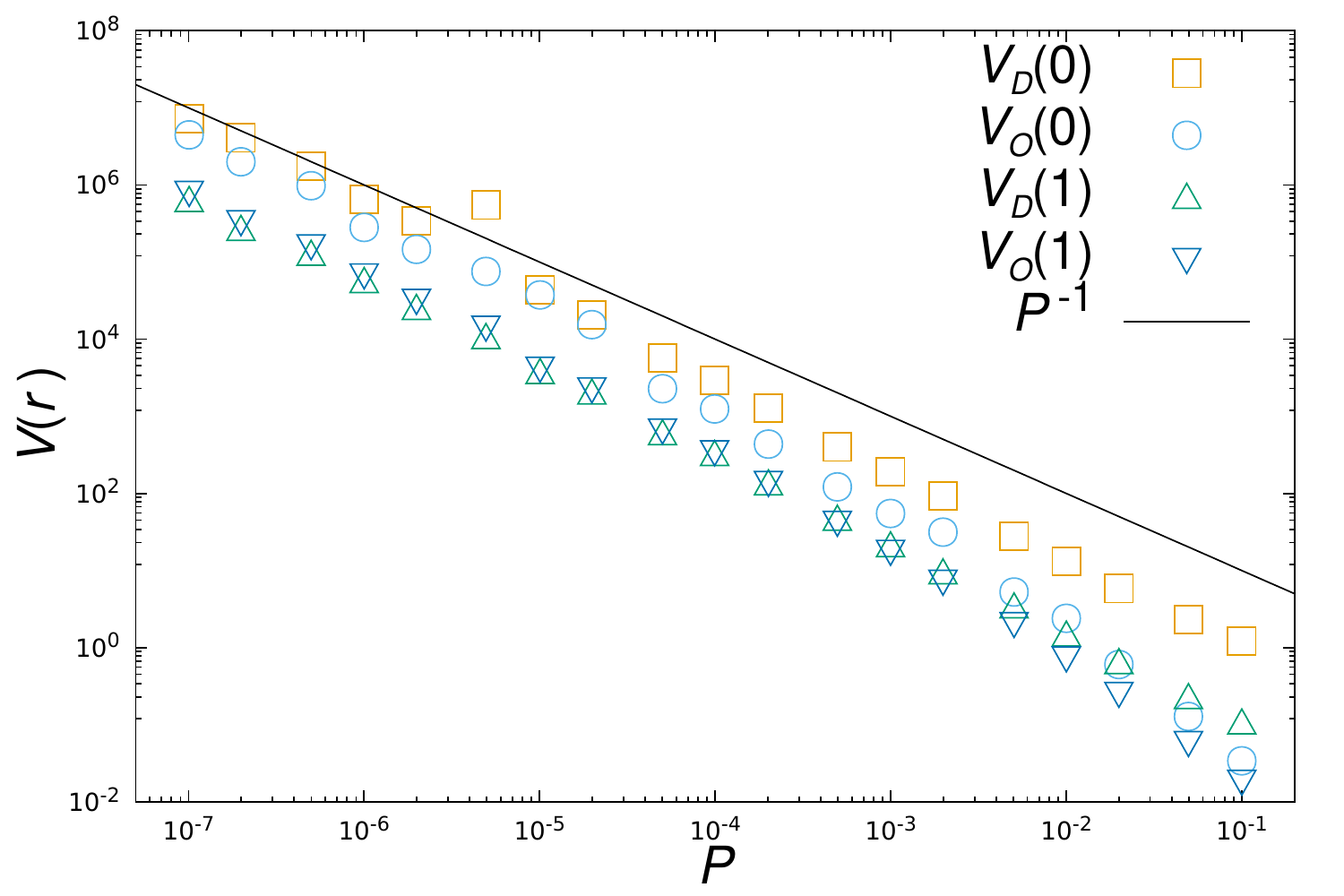}
    \caption{
    The intensities of the displacement field at $r=0$ and $r=1$ with $N=8000$ are presented.
    All of them are inversely proportional to the pressure $P$.
    Since the macroscopic shear modulus $G_{0}$ is proportional to $\sqrt{P}$ near jamming, $V(r)$ is inversely proportional to $\left(G_{0}\right)^2$, which is consistent with the prediction of elastic theory.
    %All of them are inversely proportional to pressure $P$, which means the magnitudes of $\bm{G}(0)$ and $\bm{G}(1)$ is proportional to $G_{M, 0}^{-1}$.
    }
    \label{fgr3}
\end{figure}

The results in Fig.~\ref{fgr2} clearly show that the response $r \ll l_c$ cannot be described by elastic theory at least quantitatively. 
%\blue{However, when we closely look into Fig~\ref{fgr2}~(a) in the region of $r \approx 1$, the data for different $P$ are not widely different.}
This point is important for understanding the microrheology because it measures the response of the particle on which the force is applied. 
To focus on the response on a local scale, we plot $V_D(0)$, $V_O(0)$, $V_D(1)$ and $V_O(1)$ against $P$ in Fig.~\ref{fgr3}. 
The former two are the displacement of the particle on which the point force is applied, and the latter two are those for the neighboring particles. 
Fig.~\ref{fgr3} shows that all these quantities are almost proportional to ${P}^{-1}$ near the jamming transition.
Since the macroscopic shear modulus $G_{0}$ is proportional to $\sqrt{P}$ near jamming, $V(r)$ is inversely proportional to $\left(G_{0}\right)^2$, which is the same scaling as the prediction of elastic theory. 
Thus, remarkably, these results mean that although the response of particulate packings at $r<l_c$ is heterogeneous (elastic theory breaks down), the amplitude of the displacement on a local scale is still controlled by the macroscopic shear modulus.
Since microrheology picks up response functions similar to $V_D(0), V_O(0), V_D(1), V_O(1)$, and this explains why the equivalence between micro- and macrorheology holds even though the breakdown of elastic theory takes place.

%\red{
Here we mention again that the mechanism behind the equivalence is totally different between the present jammed particles and a Newtonian fluid.
In a Newtonian fluid where continuum theory works down to the particle scale, the response on a local scale is completely described by fluid mechanics, and microrheology can measure the macroscopic viscosity.
In contrast, in jammed particles, continuum theory actually breaks down below the length scale $l_c$ and the local response cannot completely be described by elastic theory.
However, both micro- and macrorheology induce excitations of abundant anomalous modes in the plateau regime of the vDOS, which are controlled by the frequency scale $\omega_*$.
Then, the complex moduli of micro- and macrorheology are actually controlled by these anomalous modes of $\omega_*$, which is the reason why micro- and macrorheology produce equivalent results.
Note that the macroscopic shear modulus $G_0$ is controlled by $\omega_*$, as $G_0 \propto \omega_*$.
%}

\section{Concluding remarks}

In summary, we studied the microrheology of jammed particles theoretically and numerically.
We started by constructing the linear response formalism to calculate the response function $\alpha(\omega)$ of probe particles suspended in the system.
This calculation is completed by inverting the Hessian and damping matrix through the generalized eigenvalue problem. 
The obtained response function is then transformed to the complex modulus via the generalized Stokes relation.
Subsequently, we applied the developed formalism to harmonic spheres, a numerical model of jammed particles.
We have found that the complex modulus exhibits the scaling law characterized by the pressure. 
Importantly, the observed scaling law coincides with that in macrorheology even in the vicinity of the jamming point, where the characteristic length scale $l_c$ diverges.

To rationalize this equivalence, we carried out two different analyses.
First, the asymptotic analysis of the obtained formalism explains the observed scaling law from the vibrational properties of jammed particles.
Eventually, the analysis shows that the generalized Stokes relation holds due to the random and extended nature of the anomalous vibrational modes.
Second, we perform a large scale simulation to calculate the particle displacement under the point force.
Although the analysis confirms the existence of the diverging length scale $l_c$, which characterizes the heterogeneous displacement of jammed particles, we found that its magnitude near the perturbation is still controlled by the macroscopic shear modulus.

On the one hand, our results establish the validity of the microrheology of jammed particles, in the sense that it essentially reproduces the complex modulus measured in the macrorheology even near the jamming point. 
On the other hand, our results mean that microrheological measurements with a single probe particle cannot capture any trace of the diverging length scale $l_c$. 
On this point, we speculate that the length scale $l_c$ could be captured by the two-point microrheology~\cite{crocker2000, sonn2014}.
In this method, one measures correlated motions of multiple probes suspended in materials and transforms such correlated motions to the viscoelastic properties of materials assuming that the suspended medium is a continuum body.
Because the length scale $l_c$ governs the correlation between two spatially separated particles as shown in Fig.~\ref{fgr2}, the two-point microrheology has the potential to capture the presence of $l_c$ near the jamming point. 
It would be interesting to study this point in detail. 

%\blue{
Since microrheology experiments provide the broad-band complex modulus of jammed soft materials, it is quite valuable to compare our formalism with experiments.
In particular, it is interesting to experimentally extract the characteristic frequency $\omega_*$ and compare it with theory and numerical simulations at a quantitative level.
Although, in this study, we employ the simplest model for jammed soft materials, we believe that such quantitative comparison is possible when we apply our formalism to more realistic potential~\cite{lacasse1996,mason1997c}.
We are actually working in this direction.
%}

\appendix

\section{The effect of the probe size}
\begin{figure}[h]
    \includegraphics[width=0.4\linewidth]{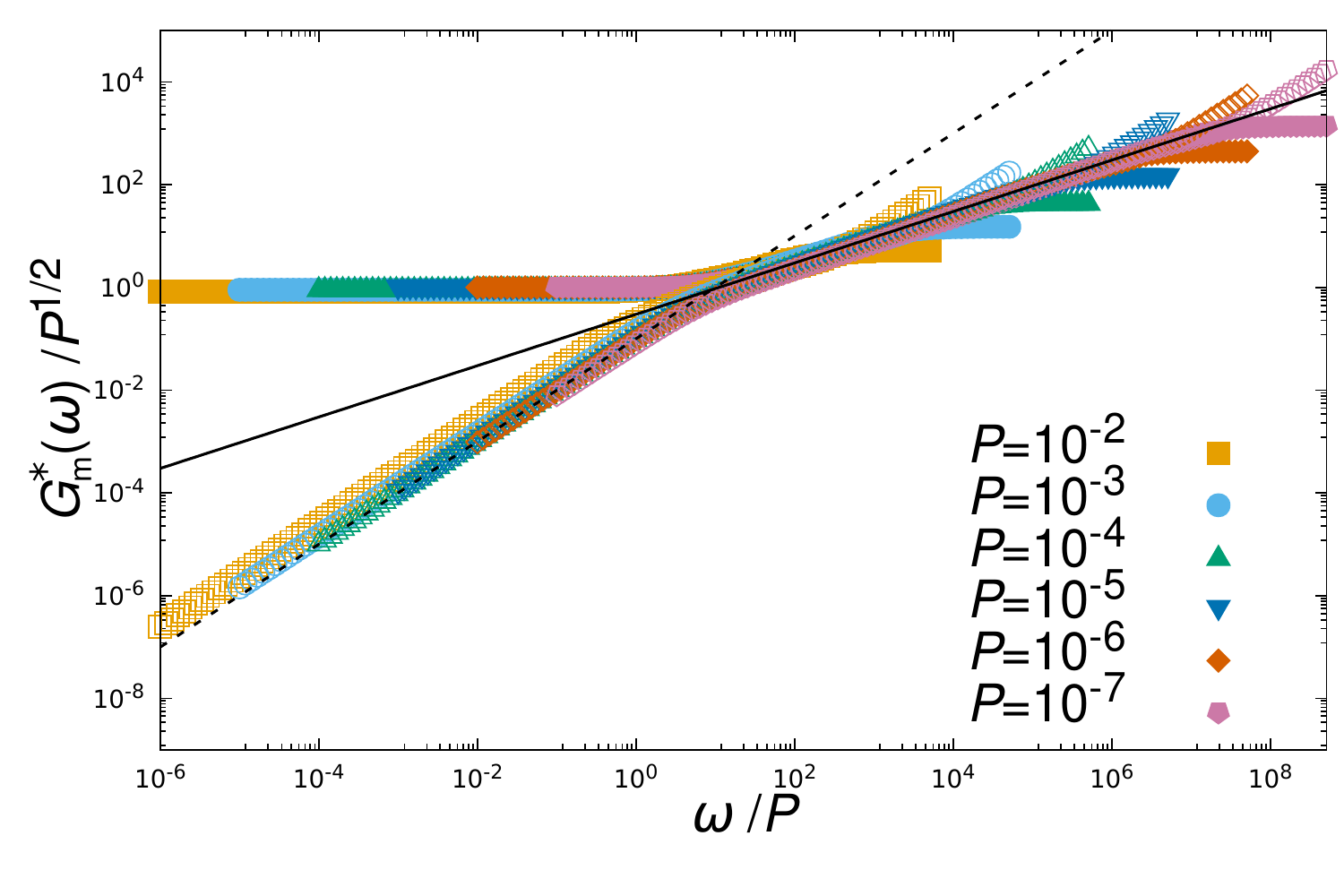}
    \caption{
    The scaling plot of the storage modulus (filled symbol) and loss modulus (open symbol) with $a_p=4$.
    This plot shows that the probe size does not affect the scaling laws obtained in $a_p=1$.
    }
    \label{fgr4}
\end{figure}

In the main text, we apply our formalism to the case where the probe particle is identical to the constituent particle.
In experiments, the size of the probe particle can be different from the constituent particles.
Thus, for completeness, we consider the case where the probe particle is larger than the constituent particle.
Here, we focus on the case of $a_p=4a, \epsilon_p=\epsilon$.
We employ the Stokes drag model with $C_{i\alpha, j\beta}=C_0\delta_{ij}\delta_{\alpha\beta}$ for constituent particles and $C_{p\alpha, p\beta}=\frac{a_p}{a}C_0 \delta_{\alpha\beta}$ for the probe particle.
The other settings such as the number of particles, the pressure range, and the reduced units are the same as in the main text.
With these settings, we calculate $G^*_m(\omega)$ by using Eq. (\ref{eq:complexmodulus}).
Fig.~\ref{fgr4} is the scaling plot of the complex modulus, and they exhibit good collapse with scaled variables.
This plot shows that the scaling laws observed with $a_p=1$ still hold with $a_p=4$.

\section{Linear response formalism for macrorheology}
Here, we summarize the linear response formalism of macrorheology \cite{lemaitre2006}.
To obtain the macroscopic complex modulus, one needs to evaluate the time evolution of the macroscopic stress induced by the strain $\eta_{\alpha\beta}(t)$.
For this purpose, first, we consider the particle motions under the time-dependent strain $\eta_{\alpha \beta}(t)$ applied at $t=0$.
With harmonic approximation, the equation of motions becomes
\begin{equation}
    \bm{C}\ket{\dot{u}} + \mathcal{M}\ket{u} = \sum_{\alpha, \beta}\eta_{\alpha\beta}(t) \ket{\Xi_{\alpha\beta}},\label{eq:shear-motion}
\end{equation}
where $\ket{\Xi_{\alpha\beta}}$ is the collection of the nonaffine force acting on the particles $\vec{\Xi}_{i, \alpha \beta} = - \pdv{\vec{f}_i}{\eta_{\alpha \beta}}$, which is induced by the strain.
With the Fourier transformation, Eq.(\ref{eq:shear-motion}) is recast as:
\begin{equation}
	\left(i \omega \bm{C} + \mathcal{M} \right) \ket{\hat{u}(\omega)} = \sum_{\alpha \beta}\hat{\eta}_{\alpha \beta}(\omega) \ket{\Xi_{\alpha \beta}},
\end{equation}
or equivalently,
\begin{equation}
    \ket{\hat{u}(\omega)} = \sum_{\alpha\beta}\hat{\eta}_{\alpha \beta}(\omega) \bm{G}(\omega)\ket{\Xi_{\alpha \beta}},
    \label{eq:fourier-shear-motion}
\end{equation}
where $\bm{G}(\omega)$ is the Green's function, as in the main text.
The particle displacement determined by Eq. (\ref{eq:shear-motion}) or (\ref{eq:fourier-shear-motion}) negatively contributes to the macroscopic stress.
The total stress increment $\Delta\sigma_{\alpha\beta}(t)$ induced by the strain is determined by the following equation:
\begin{equation}
	\Delta\sigma_{\alpha \beta}(t) = \sum_{\kappa \chi}C_{\alpha \beta \kappa \chi, \infty} \eta_{\kappa \chi}(t)  - \frac{1}{V} \ip{\Xi_{\alpha\beta}}{u},
    \label{eq:stress-evolution}
\end{equation}
where $C_{\alpha\beta\kappa\chi,\infty}$ is an affine elastic constant.
The second term is the nonaffine parts, which is the contribution from $\ket{u}$.
By the Fourier transformation, we can recast Eq. (\ref{eq:stress-evolution}) as:
\begin{equation}
	\Delta\hat{\sigma}_{\alpha \beta}(\omega) = \sum_{\kappa \chi}C_{\alpha \beta \kappa \chi, \infty} \hat{\eta}_{\kappa \chi}(\omega) - \frac{1}{V} \ip{\Xi_{\alpha\beta}}{\hat{u}(\omega)}.
    \label{eq:fourier-stress-evolution}
\end{equation}
By substituting Eq. (\ref{eq:fourier-shear-motion}) to Eq. (\ref{eq:fourier-stress-evolution}), one can obtain the following expression:
\begin{equation}
	\Delta\hat{\sigma}_{\alpha \beta}(\omega) = \sum_{\kappa \chi} \left(C_{\alpha \beta \kappa \chi, \infty} \hat{\eta}_{\kappa \chi}(\omega) - \frac{1}{V} \bra{\Xi_{\alpha \beta}} \bm{G}(\omega) \ket{\Xi_{\kappa \chi}} \hat{\eta}_{\kappa \chi}(\omega)\right).
\end{equation} 
Now, a frequency dependent elastic constant is defined as $C_{\alpha \beta \kappa \chi}(\omega) \equiv \pdv{\Delta \hat{\sigma}_{\alpha \beta}(\omega)}{\hat{\eta}_{\kappa \chi}(\omega) }$, which is explicitly written as:
\begin{equation}
	C_{\alpha \beta \kappa \chi}(\omega)
	= C_{\alpha \beta \kappa \chi, \infty} - \frac{1}{V} \bra{\Xi_{\alpha \beta}} \bm{G}(\omega) \ket{\Xi_{\kappa \chi}}.
\end{equation}
As in the main text, the Green's function is expressed by the eigenfrequencies $\omega_k$ and eigenvectors $\ket{u(\omega_k)}$ of the matrix $\tilde{\mathcal{M}}$
, and one can calculate $C_{\alpha\beta\kappa\chi}(\omega)$ as follows:
\begin{equation}
    C_{\alpha \beta \kappa \chi}(\omega)
    = C_{\alpha \beta \kappa \chi, \infty} - \frac{1}{V}
    \sum_k
    \frac{
    \bra{\Xi_{\alpha \beta}} \bm{C^{-\frac{1}{2}}} \ket{\tilde{e}(\omega_k)}
    \bra{\tilde{e}(\omega_k)}
    \bm{C^{-\frac{1}{2}}}
    \ket{\Xi_{\kappa \chi}}
    }
    {\omega_k^2 + i \omega}.
\end{equation}
The macroscopic complex modulus $G^*_M(\omega)$ presented in the main text is composed of the shear component of $C_{\alpha\beta\kappa\chi}(\omega)$ and the contribution from the solvent viscosity $\eta_{\infty}$~\cite{baumgarten2017} as follows:
\begin{equation}
    G_{M}^*(\omega) = C_{\alpha\beta\alpha\beta}(\omega) + i\omega \eta_{\infty}.
\end{equation}
In the main text, we present $G_M^*(\omega)$ averaged over the three independent shear components $xyxy, yzyz, zxzx$.

In previous studies \cite{tighe2011,baumgarten2017}, an alternative representation of the macroscopic complex modulus was proposed \cite{tighe2011,baumgarten2017}, 
\begin{equation}
    \frac{1}{G_M^*(\omega)} = V^{1-2/d}\sum_{n} \frac{\abs{\ip{\hat{\gamma}}{s_n}}^2}{\left(i\omega + s_n\right)},
    \label{eq:shear-relaxation}
\end{equation}
where, $s_n, \ket{s_n}$ are solutions of the following generalized eigenvalue problem,
\begin{equation}
    s_n\bm{B} \ket{s_n} = \tilde{\mathcal{H}} \ket{s_n},
\end{equation}
and $\ip{\hat{\gamma}}{s_n}$ is a shear component of $\ket{s_n}$ \cite{tighe2011}, which is explained later.
Here, $\tilde{\mathcal{H}}$ is the extended Hessian defined as
\begin{equation}
    \tilde{\mathcal{H}}
    =
    \left(
    \begin{array}{c|c}
    \mathcal{H} & - V^{-1/d}\ket{\Xi_{\alpha\beta}} \\ \hline 
    -V^{-1/d}\bra{\Xi_{\alpha\beta}} & V^{1-2/d}C_{\alpha\beta\alpha\beta,\infty}
    \end{array}
    \right),
\end{equation}
and $\bm{B}$ is an extended damping matrix, which becomes a diagonal matrix in the Stokes drag model as follows:
\begin{equation}
    \bm{B}
    =
    \left(
    \begin{array}{c|c}
    \bm{C} & \ket{0} \\ \hline 
    \bra{0} & V^{1-2/d} \eta_{\infty}
    \end{array}
    \right),
\end{equation}
where $\ket{0}$ is a $dN$ dimensional zero vector and $\bra{0}$ is its transpose.
Note that since $\tilde{\mathcal{H}}$ and $\bm{B}$ have an additional degree of freedom in the shear direction, $\ket{s_n}$ also has an extra component, and $\ip{\hat{\gamma}}{s_n}$ is a projection of $\ket{s_n}$ in this shear direction.

In the scaling analysis of $G^*_M(\omega)$, one assumes that the overlap with shear $\abs{\ip{\hat{\gamma}}{s_n}}^2$ does not depend on $s_n$ \cite{tighe2011}, and $\abs{\ip{\hat{\gamma}}{s_n}}^2 \approx 1/\left(dN V^{1-2/d}\right) \sim V^{2/d-2}$.
This assumption is justified by the disordered and extended nature of the mode $\ket{s_n}$.
Then, as in the same manner with microrheology, one can evaluate Eq.(\ref{eq:shear-relaxation}) as,
\begin{equation}
    \frac{1}{G^*_M(\omega)} \sim \frac{1}{V} \int d\omega^{\prime} \frac{\tilde{D}(\omega^{\prime})}{i\omega + \left(\omega^{\prime}\right)^2},
\end{equation}
where $\tilde{D}(\omega)$ is the distribution of $\sqrt{s_n}$.
Since the extended Hessian $\tilde{\mathcal{H}}$ differs from the Hessian $\mathcal{H}$ only in having an extra degree of freedom, $\tilde{D}(\omega)$ is almost identical to the vibrational density of states $D(\omega)$.
Thus, one can obtain the expression of $G_M^*(\omega)$ by $D(\omega)$, which leads to the equivalent asymptotic form of the complex modulus with microrheology discussed in the main text.

\section{Numerical calculation of the particle displacement}
Here, we summarize the numerical method to obtain the displacement field $\ket{u_0}$ of jammed particles under a force field $\ket{F}$.
Within the harmonic approximation, this is determined by the Hessian,
\begin{equation}
	\mathcal{M} \ket{u_0} = \ket{F}
        \label{eq:force-field}.
\end{equation}
Since we want to obtain $\ket{u_0}$ in a large system $(N=2048000)$, where the direct diagonalization of the Hessian is numerically hard, we introduce the following procedure.
First, we introduce the following function $E$.
\begin{equation}
	E = \frac{1}{2} \bra{u} \mathcal{M} \ket{u} + \ip{u}{F}.
\end{equation}
This function takes its minimum when $\ket{u}=\ket{u_0}$.
Thus, one can calculate $\ket{u}_0$ by minimizing $E$ with respect to $\ket{u}$.
We perform this minimization procedure by using the FIRE algorithm \cite{bitzek2006}.

In the case of the point force applied on the $i$th particle, we set $\ket{F} = ( - \frac{\va*{f}_i}{N-1}, \cdots, \va*{f}_i, \cdots,  - \frac{\va*{f}_i}{N-1})$ where $\abs{\vec{f}}_i=f_0$.
Here, the force applied to the other particles is for the conservation of the momentum of the entire system.
With the obtained displacement field, one can calculate the component of the Green's function $G_{i\alpha, j\beta}$,%from the response from $\va*{f}_j = f_0 \vu*{e}_{\beta}$ as follows:
\begin{equation}
	G_{i\alpha,j\beta} = \frac{u_{i\alpha : j\beta}}{f_{0}},
\end{equation}
where $u_{i\alpha:j\beta}$ is the displacement of particle $i$ in direction $\alpha$ under the point force on particle $j$ in direction $\beta$.

\section{The Green's function of an isotropic elastic medium}
To compare with the displacement field of jammed particles, here we summarize the prediction of elastic theory on the displacement field $\vec{u}(\vec{r})$ induced by the force field $\vec{f}(\vec{r})$.
This is determined by the following equation:
\begin{equation}
    (\mu+\lambda) \vec{\nabla} \left(\vec{\nabla} \cdot \vec{u}\right) + \mu \left(\vec{\nabla} \cdot \vec{\nabla}\right) \vec{u} + \vec{f}=\vec{0},
\end{equation}
where $\lambda, \mu$ are Lam\'{e} constants.
This is derived from the force balance condition imposed on a stress tensor and the constitutive equation of an idealized elastic body.
One can solve this equation by using the Green's function as
\begin{equation}
    \vec{u}(\vec{r}) = \int \dd \vec{r}_0 \bm{G}(\vec{r} - \vec{r}_0) \vec{f}(\vec{r}_0).
\end{equation}
The Green's function is obtained by solving the Poisson's equation, and the explicit form becomes
\begin{align}
    \bm{G}(\vec{r})
	= \frac{1}{4 \pi r \mu} \mqty(1 - \frac{1}{2} A + \frac{1}{2} A \hat{r}_x^2  &  \frac{1}{2} A \hat{r}_x \hat{r}_y & \frac{1}{2} A \hat{r}_x \hat{r}_z \\  \frac{1}{2} A \hat{r}_y \hat{r}_x & 1 - \frac{1}{2} A + \frac{1}{2} A \hat{r}_y^2 & \frac{1}{2} A \hat{r}_y \hat{r}_z \\ \frac{1}{2} A \hat{r}_z \hat{r}_x &  \frac{1}{2} A \hat{r}_z \hat{r}_y  & 1 - \frac{1}{2} A + \frac{1}{2} A \hat{r}_z^2 ),
\end{align}
where $A = \frac{\mu + \lambda}{2\mu + \lambda}$.

Now, we can calculate the intensity of the displacement field as:
\begin{align*}
	V_{D, \text{el}}(r) & = \expval{\sum_{\alpha} \left(\frac{1}{4\pi r \mu} \right)^2  \left(1 - \frac{1}{2}A + \frac{1}{2} A \hat{r}_{\alpha}^2 \right)^2}_{r} \\
	V_{O, \text{el}}(r) & = \expval{\sum_{\alpha \neq \beta} \left(\frac{1}{4\pi r \mu} \right)^2  \left(1 - \frac{1}{2}A + \frac{1}{2} A \hat{r}_{\alpha} \hat{r}_{\beta} \right)^2}_{r},
\end{align*}
where $\expval{\cdots}_r$ represents the spherical average with fixed distance $r$.
After taking the average, $V_{D, \text{el}}(r)$ and $V_{O, \text{el}}(r)$ become
\begin{align*}
	V_{D, \text{el}}(r) & = \frac{3}{5} \left(\frac{A}{8\pi r \mu} \right)^2 + 2 \left(\frac{1}{4\pi r \mu}\right)^2 \left(1 - \frac{1}{2}A \right) + \left(\frac{1}{4\pi r \mu} \right)^2 \\
	& = \left(\frac{1}{4 \pi r \mu}\right)^2 \left(\frac{3 A^2}{20} - A + 3\right) \\
	V_{O, \text{el}}(r) & = \frac{2}{5} \left(\frac{A}{8\pi r \mu} \right)^2  \\
	& = \left(\frac{1}{4\pi r \mu}\right)^2 \left(\frac{1}{10} A^2 \right).
\end{align*}

\section{Diagonal part of the intensity of the displacement}

\begin{figure}[h]
    \centering
    
    \includegraphics[width=0.5\linewidth]{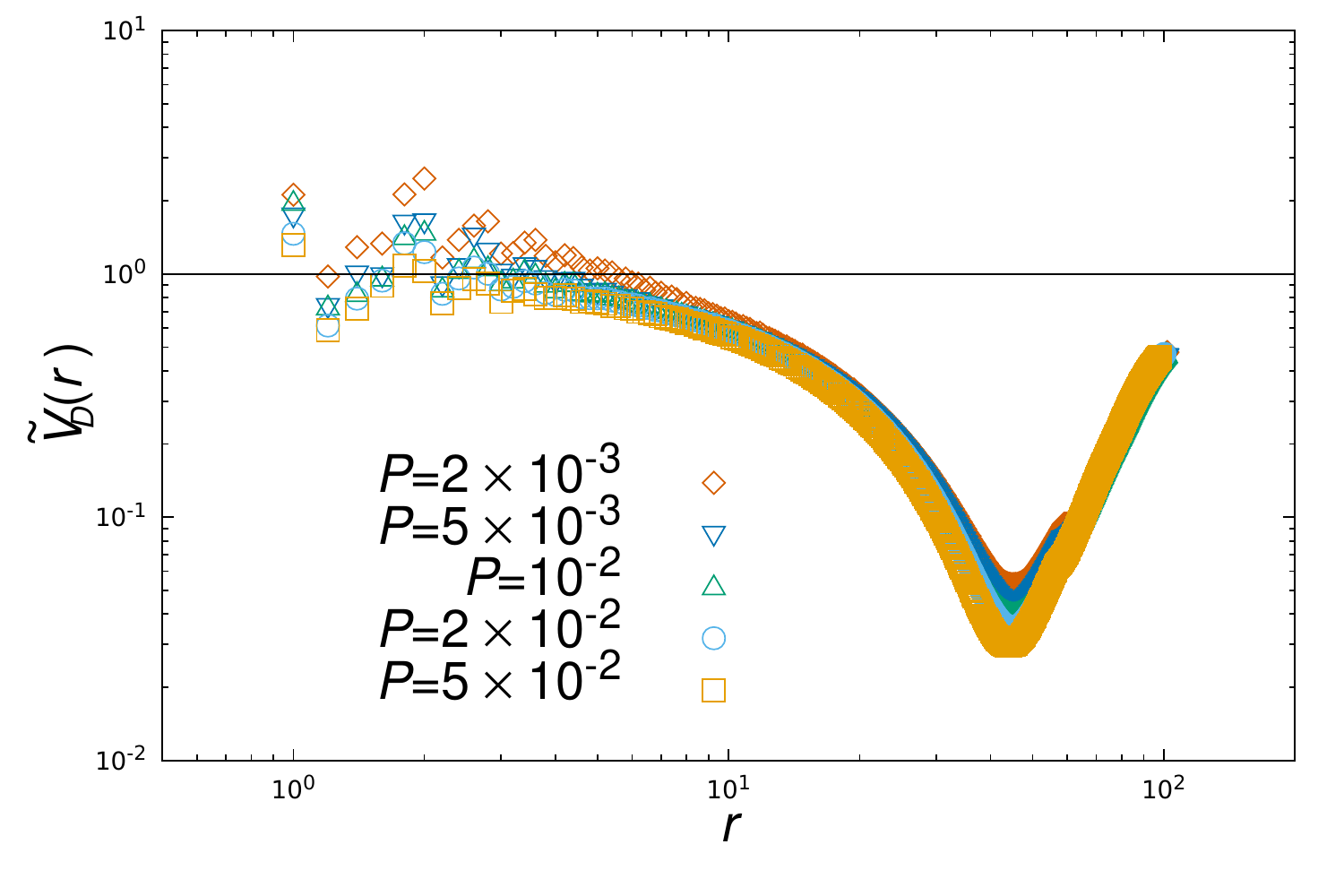}

    \caption{
    The diagonal part of the intensity of the displacement field is presented with normalization by the prediction of elastic theory.
    The diagonal part does not depend on the pressure.
    }
    \label{fgr5}
\end{figure}

In Fig.~\ref{fgr5}, we plot the diagonal part of the intensity of the displacement field.
Contrary to the off-diagonal part, the diagonal part does not exhibit pressure dependence.
We speculate that this is due to the finite system size, and the diagonal parts are very sensitive to the boundary of the system.
To clarify the reason, we need to perform calculations in larger systems, which is beyond the scope of this study.

\section*{Acknowledgements}
%The Acknowledgements come at the end of an article after Conflicts of interest and before the Notes and references.
This study was supported by Hosokawa Powder Technology Foundation (Grant Number HPTF21509), JST SPRING (Grant Number JPMJSP2108), and JSPS KAKENHI (Grant Numbers 20H00128, 20H01868, 22K03543).

\bibliography{rsc.bib} %You need to replace "rsc" on this line with the name of your .bib file
\bibliographystyle{rsc} %the RSC's .bst file

\end{document}